\documentclass[]{aa}
\usepackage{psfig}
\usepackage{txfonts}
\begin{document}

\title{Simulating star formation in molecular cores II.  The effects 
of different levels of turbulence}
\titlerunning{Simulating star formation in molecular cores II}

\author{Simon\,P.\,Goodwin, A.\,P.\,Whitworth \and D.\,Ward-Thompson}
\authorrunning{S.\,P.\,Goodwin, A.\,P.\,Whitworth \and D.\,Ward-Thompson}

\offprints{Simon.Goodwin@astro.cf.ac.uk}

\institute{Dept. of Physics \& Astronomy, Cardiff University, 5 The 
Parade, Cardiff, CF24 3YB, UK}

\date{7/5/03}

\abstract{

We explore, by means of a large ensemble of SPH simulations, how the level 
of turbulence affects the collapse and fragmentation of a star-forming core. 
All our simulated cores have the same mass ($5.4 M_\odot$), 
the same initial density profile (chosen to fit observations of L1544), 
and the same barotropic equation of state, but we vary (a) the initial 
level of turbulence (as measured by the ratio of turbulent to gravitational 
energy, $\alpha_{\rm turb} \equiv U_{\rm turb}/|\Omega| = 0,\,0.01,\,0.025,\,
0.05,\,0.10\;{\rm and}\;0.25$) and (b), for fixed $\alpha_{\rm turb}$, the 
details of the initial turbulent velocity field (so as to obtain good 
statistics).

\hspace{0.3cm} A low level of turbulence ($\alpha_{\rm turb} \sim 0.05$) 
suffices to produce multiple systems, and as $\alpha_{\rm turb}$ is increased, 
the number of objects formed and the companion frequency both increase. The mass 
function is bimodal, with a flat low-mass segment representing single objects 
ejected from the core before they can accrete much, and a Gaussian high-mass 
segment representing objects which because they remain in the core grow 
by accretion and tend to pair up in multiple systems.

\hspace{0.3cm}The binary statistics reported for field G-dwarfs by Duquennoy 
\& Mayor (1991) are only reproduced with $\alpha_{\rm turb} \sim 0.05$. For 
much lower values of $\alpha_{\rm turb}$ ($\la 0.025$), insufficient binaries 
are formed. For higher values of $\alpha_{\rm turb}$ ($\ga 0.10$), there
is a significant sub-population of binaries with small semi-major axis 
and large mass-ratio (i.e. close binaries with 
components of comparable mass). This sub-population is not present in 
Duquennoy \& Mayor's sample, although there is some evidence for it in 
the pre-Main Sequence population of Taurus analyzed by White \& Ghez (2001). 
It arises because with larger $\alpha_{\rm turb}$, more low-mass objects 
are formed, and so there is more scope for the binaries remaining in the 
core to be hardened by ejecting these low-mass objects. Hard binaries thus 
formed then tend to grow towards comparable mass by competitive accretion 
of material with relatively high specific angular momentum.

}

\maketitle

\section{Introduction}

Turbulence appears to play a crucial role in the 
structure and evolution of molecular clouds, in the formation 
of star-forming cores within molecular clouds, and in the 
collapse and fragmentation of cores to form protostars.

The main evidence for turbulence in molecular clouds comes from 
their apparently fractal substructure (e.g. Elmegreen \& Falgarone 
1996; Elmegreen 2002), and from the almost universal power-law 
scaling relations between size ($L$) line-width ($\Delta v$) and 
mass ($M$)\footnote{$L \propto M^\alpha$ with $0.35 \la \alpha \la 
0.55\,$, and $\Delta v \propto M^\beta$ with $0.2 \la \beta \la 0.3$} 
and the almost universal power-law mass function\footnote{$d{\cal N}/dM 
\propto M^{-\gamma}$ with $1.5 \la \gamma \la 2.0$} 
to which this substructure subscribes, over many orders of magnitude, 
from the largest molecular cloud complexes ($\sim\!10^2\,{\rm pc}$, 
$\,\sim\!10^6 M_\odot$), down to the smallest resolvable structures 
($\sim\!10^{-2}\,{\rm pc}\,$, $\,\sim\!10^{-4} M_\odot$) (e.g. Larson 1981; 
Myers 1983; Stutzki \& G\"usten 1990;  Hobson 1992; Hobson et al. 1994; 
Williams et al. 1994; Elmegreen \& Falgarone 1996; Kramer et al. 1996; 
Kramer et al. 1998; Heithausen et al. 1998).

Until recently, it had been presumed that molecular clouds were long 
lived, being supported against collapse by their internal turbulence, 
and 
this was advanced as the reason for the low 
overall efficiency of star formation.. However, it is 
now recognized that turbulence cannot support clouds for long, 
because -- even with a frozen-in magnetic field -- the turbulence 
dissipates on a dynamical timescale (Mac Low et al. 1998; 
Stone et al. 1998). Instead clouds are presumed to 
form and disperse on a dynamical timescale, without ever reaching 
equilibrium (Ballesteros-Paredes et al. 1999; Elmegreen 2000; 
Pringle et al. 2001; Hartmann et al. 2001).

In this highly dynamical scenario, cores form wherever a 
sufficiently dense and coherent converging flow is created by 
the turbulent velocity field (Elmegreen 1997; Padoan et al. 
1997; Hartmann et al. 2001; Klessen \& Burkert 2000, 2001; 
Klessen et al. 2000; Padoan \& Nordlund 2002; Mac Low \& Klessen 
2004). Frequently these cores are not gravitationally bound, and 
therefore they disperse soon after they form. However, occasionally they are 
gravitationally bound, and in this case they are likely to proceed 
straight into gravitational collapse; these are the cores we identify 
as `prestellar'.

This scenario is in contrast to the idea that prestellar cores 
are supported magnetically, and evolve quasistatically by 
ambipolar diffusion, until they become magnetically supercritical 
and collapse (e.g. Basu \& Mouschovias 1994, 1995a,b; Ciolek \& 
Mouschovias 1993, 1994, 1995; Morton et al. 1994; Ciolek \& Basu 2000). 
The main effects of the quasistatic ambipolar diffusion phase are (i) 
to give the core more time to lose angular momentum by magnetic braking, 
(ii) to organize the material so that its subsequent collapse is 
rather well focussed, and (iii) to allow turbulence to decay. All
three effects mean that such cores are less likely to form multiple 
systems. Since most stars are observed to be in multiple systems
(e.g. Duquennoy \& Mayor 1991; Fischer \& Marcy 1992; White \& Ghez 
2001), and since there is no observational evidence 
for magnetically subcritical cores (e.g. Crutcher 1999; Bourke et al. 
2001; Crutcher at el. 2003), we shall assume that ambipolar diffusion 
does not play an important role in the evolution of prestellar cores. 
For simplicity, we ignore the magnetic field altogether.

In the highly dynamic scenario the collapse of a prestellar core is 
far more likely to lead to fragmentation and the formation of multiple 
systems (e.g. Whitworth et al. 1995; Turner et al. 1995; Whitworth et al. 
1996; Klein et al. 2001, 2003; Bate et al. 2002a,b, 2003; Bonnell et al. 
2003; Delgado-Donate 2003, 2004; Goodwin et al. 2004a,b; Hennebelle et al. 
2003, 2004). This is because in the highly dynamic scenario prestellar 
cores are formed non-quasistatically and therefore (a) they are launched 
directly into the non-linear regime of gravitational collapse, and (b) 
they are likely to have retained some internal turbulence.  

Burkert \& Bodenheimer (2000) have pointed out that the internal turbulence 
in molecular cores can be represented by a Gaussian random velocity field 
having a power spectrum of the form $P(k) \propto k^\eta$, with 
$\eta \sim -3\;{\rm to}\;-4$. This not only reproduces the observed 
scaling relation between size and linewidth. It also reproduces the 
observed rotation of molecular cores. Thus there is no need to invoke 
ordered rotation as an additional source of support in molecular cores, 
and indeed there is no observational evidence for significant ordered 
rotation in prestellar cores (e.g. Jessop \& Ward-Thompson 2001).

We have therefore undertaken a numerical study of the influence of 
turbulence on the collapse and fragmentation of prestellar cores. We 
have taken as our reference point a simple model of the core L1544,
and in the first paper of this series (Goodwin et al. 2004a, hereafter
Paper I) we have shown that cores with even a low level of turbulent 
energy routinely spawn multiple stellar systems. Specifically, in an 
ensemble of 20 simulations of the collapse of $5.4 M_\odot$ cores 
having an initial ratio of turbulent to gravitational energy
\begin{equation}
\label{ALPHATURB}
\alpha_{\rm turb} \equiv \frac{U_{\rm turb}}{|\Omega|} = 0.05 \,,
\end{equation}
$80\%$ of the cores form at least two, and in one case ten, objects (stars 
and brown dwarfs). In addition, the distributions of semi-major axis, 
mass ratio and eccentricity for the resulting multiple systems are 
consistent with the distributions for observed binary systems (e.g. 
Duquennoy \& Mayor 1991, hereafter DM91). Paper I has also shown that while 
low levels of turbulence can easily produce multiple fragmentation, the 
outcome of any one simulation depends sensitively on the exact details 
of the turbulent velocity field. Consequently a statistical approach is 
essential in evaluating the influence of turbulence on multiple star 
formation in cores.

In this paper we extend the simulations of Paper I to examine the
effect of different levels of turbulence on star formation within 
dense molecular cores.  Using exactly the same core structure 
as in Paper I we simulate ensembles of between 10 and 20 cores with 
$\alpha_{\rm turb} = 0,\,0.01,\,0.025,\,0.05,\,0.10\;{\rm and}\;0.25$. 
We examine the numbers and masses of stars and brown dwarfs that 
form and the properties of the multiple systems to which some of them 
belong.

We note that these levels of turbulence involve much lower non-thermal 
velocities than the earlier work of Whitworth et al. (1995), Turner et
al. (1995), Whitworth et al. (1996), Klein et al. (2001, 2003), Bate
et al. (2002a,b, 2003), Bonnell et al. (2003), Delgado-Donate (2003, 
2004) Goodwin et al. (2004a,b) Hennebelle et al. , and Hennebelle et 
al. (2003, 2004). Consequently they may be applicable to scenarios 
in which instability develops more quasistatically due to ambipolar 
diffusion. provided that some turbulence can persist through (or 
be regenerated after) the ambipolar diffusion phase, and provided the 
subsequent collapse is sufficiently rapid.

In Section 2 we define the initial conditions for the simulations. 
In Section 3 we describe the code and the constitutive physics used. 
In Section 4 we outline the different ensembles of simulations 
performed with different values of $\alpha_{\rm turb}$, and in 
Section 5 we collate the statistics from the different ensembles. 
Section 6 discusses the statistics in terms of the underlying physics, 
and Section 7 gives our main conclusions.

\section{Initial conditions}

Molecular cores which are associated with IRAS sources are presumed 
to have already formed a protostar, and are classified as 
protostellar cores, whereas those which have no associated IRAS source 
are classified as starless cores (Beichman et al. 1986). The densest 
starless cores are presumed to be on their way to forming stars, and are 
therefore classified as prestellar cores (Ward-Thompson et al. 1994). 
We base our initial conditions on the observed properties of prestellar 
cores.

The density in a pre-stellar core is approximately uniform in the inner 
few thousand au, but further out it decreases as $r^{-\eta}$ with 
$2 \la \eta \la 5$, and eventually it merges with the background (e.g. 
Ward-Thompson et al. 1994; Andr\'e et al. 1996; Ward-Thompson et al. 
1999; Andr\'e et al. 2000; Tafalla et al. 2004). A good fit to the 
density in a pre-stellar core is given by a Plummer-like profile,
\begin{equation}
\rho(r) = \frac{\rho_{\rm kernel}}{(1 + (r/R_{\rm kernel})^2)^2} \,,
\end{equation}
where $\rho_{\rm kernel}$ is the central density and $R_{\rm
kernel}$ is the radius of the region in which the density is approximately 
uniform. We set $\rho_{\rm kernel} = 3 \times 10^{-18}\,{\rm g}\,{\rm cm}^{-3}$ 
and $R_{\rm kernel} = 5 000\,{\rm au}$, with the outer boundary of the core 
at $R_{\rm core} = 50 000\,{\rm au}$, so the core has total mass 
$M_{\rm core} = 5.4 M_{\odot}$. The core is initially isothermal with  
$T = 10\,{\rm K}$, and uniformly molecular with mean gas-particle mass 
$\bar{m} = 4 \times 10^{-24}\,{\rm g}$. This means that the core has a 
ratio of thermal to gravitational energy of 
\begin{equation}
\alpha_{\rm therm} \equiv \frac{U_{\rm therm}}{|\Omega|} \simeq 0.3 \;.
\end{equation}
Thus far, these are exactly the same initial conditions as we used in Paper I.

\begin{figure*}
\centerline{\psfig{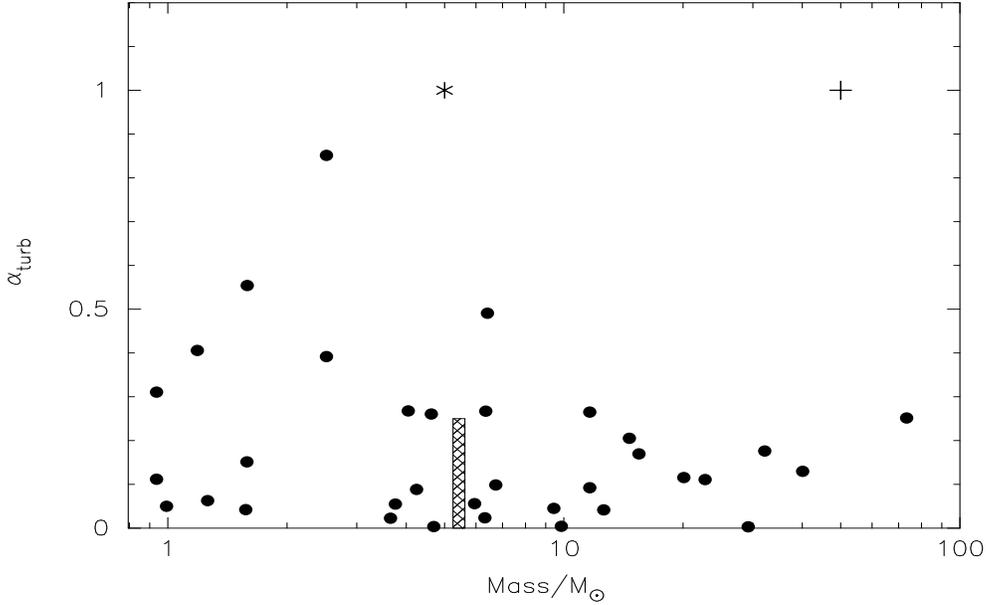}}
\caption{The filled circles give estimated values of $\alpha_{\rm turb}$ 
(the ratio of turbulent to gravitational energy, see Eqn. 
(\ref{ALPHATURB})) and $M_{\rm core}$ (core 
mass) for the starless cores in the Jijina et al. (1999) catalogue. The 
hatched region gives the range of $(\alpha_{\rm turb},M_{\rm core})$ values 
used in this paper (i.e. $0 \leq \alpha_{\rm turb} \leq 0.25$ and 
$M_{\rm core} = 5.4 M_\odot$). The cross ($\times$) and the star ($*$) 
give -- respectively -- the $(\alpha_{\rm turb},M_{\rm core})$ values 
used by Bate et al. (2002a,b;2003) and Delgado Donate et al. (2003a,b).}   
\label{fig:nonthermal}
\end{figure*}

Molecular cores show a significant non-thermal contribution to their 
line-widths, which is attributable to internal turbulence. 
Fig.~\ref{fig:nonthermal} shows the estimated ratios of turbulent to 
gravitational energy, $\alpha_{\rm turb}$ (see Eqn. (\ref{ALPHATURB})), 
and the estimated masses, $M_{\rm core}$, for prestellar cores 
from the Jijina et al. (1999) catalogue. These cores were selected 
as prestellar on the basis of their having low temperature ($<20\,{\rm K}$), 
no associated IRAS source and no observed outflow. The shaded area 
in Fig.~\ref{fig:nonthermal} shows the region of parameter space that 
the simulations in this paper cover, i.e. a $5.4 M_{\odot}$ core with 
a range of $\alpha_{\rm turb}$ from 0 to 0.25.

To model the turbulence, a divergence-free gaussian random velocity 
field is superimposed on the core (cf. Bate et al. 2002a,b; Bate et al. 
2003; Fisher 2004; Bonnell et al. 2003; Delgado-Donate et al. 2003, 2004). 
The power spectrum of the velocity field is $P(k) \propto k^{-4}$, so 
as to mimic the observed scaling between size and line-width in 
interstellar gas clouds (Larson 1981; Burkert \& Bodenheimer 2000). 
The magnitude of the velocity field is normalized so that $\alpha_{\rm 
turb} = 0,\,0.01,\,0.025,\,0.05,\,0.1\;{\rm or}\; 0.25\,$, and at 
least ten realizations have been made with each value of $\alpha_{\rm turb}$ 
(as summarised in Table~\ref{tab:runs}).

In each realization the random number seed for the turbulence is different, 
and hence the detailed structure of the velocity field is different. It is 
essential that many different realizations be performed for a given value of 
$\alpha_{\rm turb}$, because the mix of protostars and multiple 
systems which forms depends critically on the details of the velocity field. 
Therefore different values of $\alpha_{\rm turb}$ can only be compared in a 
statistical sense by performing an ensemble of different realizations for 
each representative value of $\alpha_{\rm turb}$.

Each simulation is evolved for $0.30\,{\rm Myr}$. We have chosen this 
end-time for three reasons. (i) It facilitates comparison with the 
$\alpha_{\rm turb} = 0.05$ simulations reported in Paper I. (ii) In 
most simulations, object formation (i.e. sink creation) ceases around 
$0.15\,{\rm to}\,0.20\,{\rm Myr}$, and in only one case is a new object formed 
after $0.25\,{\rm Myr}$. (iii) By this stage, $\sim 66\,\%$ of the mass 
is already in stars and brown dwarves (this fraction decreases slightly 
with increasing $\alpha_{\rm turb}$), and so the remaining diffuse gas is 
likely to be affected by negative feedback from the existing stars and 
brown dwarves; feedback is not included in these simulations, although 
we are currently exploring its effect (Boyd et al. in preparation).

The properties of the resulting protostars in each ensemble are 
compared, {\it both} as a function of 
$\alpha_{\rm turb}$, {\it and} against the observed properties of young 
stellar objects. In this latter regard, the observational samples used for 
comparison are the local field G-dwarfs studied by DM91 and the 
multiple systems in Taurus studied by White \& Ghez (2001; hereafter WG01).

\section{Computational method and constitutive physics}

The simulations are performed using SPH (Lucy 1977; Gingold \& 
Monaghan 1977; Monaghan 1992). Our SPH code ({\sc dragon}) uses 
the standard M4 kernel (Monaghan \& Lattanzio 1985) and invokes 
variable smoothing lengths (so that each particle has 
${\cal N}_{\rm neib} = 50 \pm 5$ neighbours). An octal tree is 
built to facilitate the computation of gravitational accelerations 
and the identification of neighbours. Gravity is kernel-softened 
with the particle smoothing length, and standard artificial viscosity 
(Monaghan 1992) is included, with $\alpha_v=1$ and $\beta_v=2$.

We use a barotropic equation of state, 
\begin{equation}
\label{EofS}
\frac{P(\rho)}{\rho} \equiv c_s^2(\rho) = c_0^2 \left[ 1 + \left
( \frac{\rho}{\rho_{\rm crit}} \right)^{2/3} \right] \,.
\end{equation}
Thus at low densities ($\rho < \rho_{\rm crit} = 
10^{-13}\,{\rm g}\,{\rm cm}^{-3}$, where radiative cooling is optically 
thin) the gas is approximately isothermal with $c_s \simeq 
c_0 = 0.2\,{\rm km}\,{\rm s}^{-1}$, corresponding to molecular gas 
at temperature $T = 10\,{\rm K}$; and at high densities, ($\rho > 
\rho_{\rm crit}$, where cooling radiation is trapped by the 
high optical-depth) the gas behaves adiabatically, with ratio of 
specific heats $\gamma \simeq 5/3$. ($\gamma \simeq 5/3$ is the 
appropriate ratio of specific heats because the rotational degrees 
of freedom of H$_2$ are not excited until the temperature rises 
above $\sim 300\,{\rm K}$, and our simulations do not follow the 
gas dynamics to such high temperatures; see below). Eqn. (\ref{EofS}) 
reproduces the thermal behaviour of the gas in detailed one-dimensional 
simulations of the collapse of $1 M_\odot$ cores (e.g. Larson 1969; 
Tohline 1982; Masunaga \& Inutsuka 2000).

Whenever a gravitationally bound condensation forms and the density 
of an SPH particle within the condensation rises above $100 \rho_{\rm crit} 
= 10^{-11}\,{\rm g}\,{\rm cm}^{-3}$, all the particles within $5\,{\rm au}$ 
of that particle are replaced with a sink particle having radius $R_{\rm sink} 
= 5\,{\rm au}$. (From Eqn. (\ref{EofS}), we can estimate that the temperature 
of gas at $\rho = 100 \rho_{\rm crit}$ is $\sim 225\,{\rm K}$, and so 
the introduction of sink particles makes it unnecessary to treat the 
thermal behavious of the gas at temperatures above $\sim 225\,{\rm K}$.) 
Sink particles interact with the gas both gravitationally, and by 
accreting SPH particles that enter the sink radius and are bound to 
the sink (see Bate et al. 1995 for a detailed description of sink 
particles). As in Paper I, we refer to sink particles as `objects', 
and then more specifically to `stars' when the sink mass is $> 0.08 
M_{\odot}$ and `brown dwarfs' when the mass is lower than this.

\section{Ensembles of simulations with different $\alpha_{\rm turb}$}

\subsection{$\alpha_{\rm turb} = 0$ (no turbulence)}

As reported in Paper I, when a core has no turbulence only one object
forms, very close to the centre of the core.  The evolution follows 
very closely the semi-analytic model of Whitworth \& Ward-Thompson 
(2001). In particular, the accretion rate is very large early on, and 
then decreases. After $0.3\,{\rm Myr}$ the stellar mass reaches $3.75 
M_{\odot}$. (In this case the ten different realizations involve 
different initial SPH particle positions.)

\subsection{$\alpha_{\rm turb} = 0.01$}

The ensemble of ten simulations with $\alpha_{\rm turb} = 0.01$ is 
virtually identical to that with $\alpha_{\rm turb} = 0.00$ (no 
turbulence). Only one star ever forms, and this happens about one 
free-fall time ($\sim 0.055\,{\rm Myr}$) after the start of the 
simulation, always close to the centre of mass of the core. This 
level of turbulence is apparently too low to induce multiple 
fragmentation.

\subsection{$\alpha_{\rm turb} = 0.025$}

$\alpha_{\rm turb} = 0.025$ appears to be approximately the minimum 
level of turbulence required for multiple object formation. Of the ten 
simulations with $\alpha_{\rm turb} = 0.025$, eight produce only a 
single star (as with lower levels of turbulence), but one simulation 
produces six objects, and one produces nine objects.

Specifically, this last simulation produces three intermediate-mass stars 
in an hierarchical triple system embedded in the core (a tight binary with 
component masses $1.07 M_\odot$ and $0.88 M_\odot$ and semi-major axis 
$8.6\,{\rm au}$, plus a third star with mass $1.65 M_\odot$ orbiting at 
$\sim 70\,{\rm au}$). In addition, the simulation produces one very 
low-mass star ($0.087 M_\odot$) and five brown dwarfs ($0.034 M_\odot$ 
to $0.072 M_\odot$), all of which are ejected from the core.

\subsection{$\alpha_{\rm turb} = 0.05$}

Of the twenty simulations performed with $\alpha_{\rm turb} = 0.05$, 
four produce just a single star, and the remaining sixteen produce 
71 stars and 16 brown dwarfs between them (between 2 and 10 objects 
per simulation). Of these 71 stars, 44 
remain in the core in multiple systems, and the rest are ejected 
from the core. Of the 16 brown dwarfs, 15 are ejected, and only 
one remains in a binary system in the core. The mean number of 
objects formed per simulation is 4.55. Further details of this 
ensemble of simulations are given in Paper I.

\subsection{$\alpha_{\rm turb} = 0.10$}

Of the twenty simulations performed with $\alpha_{\rm turb} = 0.10$, 
five produce just a single star, and the remaining fifteen produce 
76 stars and 19 brown dwarfs between them (between 3 and 10 objects 
per simulation). Of these 76 stars, 53 remain in the core in multiple 
systems, and the rest are ejected from the core. Of the 19 brown dwarfs, 
17 are ejected and only 2 remain in multiple systems in the core. The 
mean number of objects formed per simulation is 4.75.

\subsection{$\alpha_{\rm turb} = 0.25$}

In the ten simulations with $\alpha_{\rm turb} = 0.25$, a total of 60
objects are produced, 49 stars and 11 brown dwarfs, with each simulation 
producing between 3 and 10 objects. Of the 49 stars, 36 remain in the 
core in multiple systems, and the rest are ejected from the core. All 
11 brown dwarfs are ejected from the core. 

\begin{table*}
\caption[]{For each value of $\alpha_{\rm turb}$, we list the number of 
realizations simulated (${\cal N}_{\rm real}$), the mean mass in objects 
at the end of the simulations ($M_{\rm tot}/M_\odot$), the average number 
of objects formed per  simulation $\left(\left<{\cal N}_{\rm obj}\right>\right)$, 
the companion probability ({\bf cp}), the companion frequency ({\bf cf}), 
the multiplicity frequency ({\bf mf}) and the pairing factor ({\bf pf}). 
The companion frequency, {\bf cf}, is given for all objects and then 
seperately for low-mass objects ($M < 0.5 M_\odot$) and for high-mass 
objects ($M > 0.5 M_\odot$).}
\label{tab:runs}
\begin{center}
\begin{tabular}{cccccccccccc}
 & \\
\hline
 & \\
$\alpha_{\rm turb}$ & ${\cal N}_{\rm real}$ & $M_{\rm tot}/M_\odot$
 & $\left<{\cal N}_{\rm obj}\right>$ & {\bf cp} & & & {\bf cf} & & & {\bf mf} 
 & {\bf pf} \\
 & & & & & & all & low & high & & & \\
 & \\
\hline
 & \\
0.00  & 10 & 3.75 & 1 & & & & & & & & \\
0.01  & 10 & 3.75 & 1 & & & & & & \\
0.025 & 10 & 3.70 & $2.3\pm0.5$ & & & & & & \\
0.05  & 20 & 3.57 & $4.6\pm0.5$ & 0.51 & & 0.95 & 0.30 & 1.45 & & 0.27 & 1.71 \\
0.10  & 20 & 3.35 & $4.8\pm0.5$ & 0.59 & & 1.39 & 0.48 & 2.32 & & 0.32 & 2.11 \\
0.25  & 10 & 3.14 & $6.0\pm0.5$ & 0.62 & & 1.57 & 0.81 & 2.38 & & 0.32 & 2.36 \\
&\\
\hline
\end{tabular}
\end{center}
\end{table*}


\begin{table*}
\caption[]{A summary of the results of the simulations with $\alpha_{\rm turb} 
= 0.05,\,0.10\;{\rm and}\;0.25\,$, at time $t = 0.3\,{\rm Myr}\,$. 
Column 1 gives the simulation identifier and Column 2 gives 
$\alpha_{\rm turb}$. Column 3 gives $M_{\rm obj}$, the total 
mass of objects formed (stars plus brown dwarfs), Column 4 gives 
${\cal N}_{\rm obj}$, the total number of objects formed, and 
Column 5 gives ${\cal N}_{\rm BD}$, the total number of brown 
dwarfs formed. Column 6 gives the multiplicities of the multiple 
systems formed, and the final column (Column 7) gives the mass 
of each individual object. Those objects which are part of a 
binary system are distinguished with $^b$, those which are part 
of a triple system are distinguished with a $^t$, and those 
which are part of a quadruple system (or in one case a quintuple 
system) are distinguished with $^q$.}
\label{tab:bins}
\begin{tabular}{lllllll}
\hline
&&&&&\\
 {\rm ID} & $\alpha_{\rm turb}$ & $M_{\rm obj}$ & $N_{\rm obj}$ &
$N_{\rm bd}$ & Multiplicity & Masses/$M_{\odot}$ \\
&&&&&\\
\hline
&&&&&\\

071 & 0.05 & 2.94 & 7  & 2 & Triple     & 1.31$^t$, 0.61$^t$,
0.52$^t$, 0.27, 0.12, 0.063, 0.048  \\
072 & 0.05 & 3.72 & 4  & 0 & Binary     & 2.32$^b$, 0.74$^b$, 0.48,
0.18 \\
073 & 0.05 & 3.10 & 10 & 2 & Binary + Triple?  & 1.07$^b$, 0.66$^b$, 
0.43, 0.34, 0.17, 0.13$^t$, 0.10$^t$, 0.09$^t$, 0.076, 0.040 \\
074 & 0.05 & 4.02 & 3  & 0 & Triple     & 1.63$^t$, 1.56$^t$, 0.83$^t$ \\
075 & 0.05 & 3.69 & 2  & 0 & Binary     & 2.63$^b$, 1.06$^b$  \\
076 & 0.05 & 3.61 & 3  & 1 & Binary     & 2.18$^b$, 1.40$^b$, 0.028\\
077 & 0.05 & 3.75 & 6  & 1 & Triple     & 1.60$^t$, 1.16$^t$, 0.64$^t$, 
0.18, 0.12, 0.050  \\
078 & 0.05 & 3.65 & 7  & 2 & Triple     & 1.09$^t$, 1.03$^t$, 0.69$^t$, 
0.58, 0.18, 0.045, 0.041\\
079 & 0.05 & 3.81 & 8  & 3 & Triple     & 1.27$^t$, 1.16$^t$, 0.69$^t$, 
0.39, 0.21, 0.044, 0.030, 0.025\\
080 & 0.05 & 3.63 & 1  & 0 & Single     & 3.63\\
081 & 0.05 & 3.69 & 1  & 0 & Single     & 3.69  \\
082 & 0.05 & 4.01 & 4  & 0 & Quadruple  & 1.52$^q$, 0.91$^q$, 0.89$^q$, 
0.69$^q$  \\
083 & 0.05 & 3.56 & 4  & 0 & Triple     & 1.43$^t$, 0.83$^t$, 0.70$^t$, 
0.60  \\
084 & 0.05 & 3.55 & 5  & 0 & Binary     & 1.46$^b$, 1.28$^b$, 0.43, 
0.19, 0.18 \\
085 & 0.05 & 3.47 & 8  & 3 & Binary     & 1.43$^b$, 0.76$^b$, 
0.51, 0.47, 0.14, 0.064, 0.045, 0.039 \\
086 & 0.05 & 3.94 & 7  & 1 & {\rm Triple}     & 1.23$^t$, 1.03$^t$, 0.73, 
0.71$^t$, 0.11, 0.098, 0.027\\
087 & 0.05 & 3.67 & 2  & 0 & {\rm Binary}     & 3.19$^b$, 0.48$^b$ \\
088 & 0.05 & 3.35 & 1  & 0 & {\rm Single}     & 3.35 \\
089 & 0.05 & 3.61 & 7  & 1 & {\rm Quadruple}  & 1.20$^q$, 0.89$^q$, 
0.57, 0.51, 0.29$^q$, 0.11, 0.041$^q$  \\
090 & 0.05 & 2.62 & 1  & 0 & {\rm Single}     & 2.62 \\

001 & 0.10 & 3.78 & 3  & 0 & {\rm Triple}     & 1.49$^t$, 1.15$^t$,
1.13$^t$ \\
002 & 0.10 & 2.83 & 1  & 0 & {\rm Single}     & 2.38\\
003 & 0.10 & 3.72 & 1  & 0 & {\rm Single}     & 3.72 \\
004 & 0.10 & 3.48 & 1  & 0 & {\rm Single}     & 3.48 \\
005 & 0.10 & 2.86 & 4  & 1 & {\rm Binary}     & 1.43$^b$, 0.65$^b$, 0.77,
0.02 \\
006 & 0.10 & 2.84 & 1  & 0 & {\rm Single}     & 2.84 \\
007 & 0.10 & 3.15 & 5  & 0 & {\rm Triple\,\,\&\,\,Binary} & 1.76$^t$,
0.72$^t$, 0.47$^t$, 0.10$^b$, 0.10$^b$  \\
008 & 0.10 & 3.22 & 6  & 2 & {\rm Quadruple}  & 1.97$^q$, 0.47$^q$,
0.35$^q$, 0.34$^q$, 0.03, 0.06 \\
009 & 0.10 & 3.48 & 8  & 4 & {\rm Quadruple}  & 2.28$^q$, 0.49$^q$,
0.26$^q$, 0.25$^q$, 0.05, 0.08, 0.04, 0.04 \\
010 & 0.10 & 3.31 & 8  & 1 & {\rm Quadruple}  & 0.76$^q$, 0.74$^q$,
0.58$^q$, 0.57$^q$, 0.08, 0.09, 0.03, 0.46 \\
011 & 0.10 & 3.96 & 12 & 4 & {\rm Triple\,\,\&\,\,binary?} & 0.89$^t$, 
0.82$^t$, 0.82$^t$, 0.04$^b$, 0.04$^b$, 0.42, 0.38, 0.03, 0.03, 0.25,
0.12, 0.11 \\
012 & 0.10 & 3.60 & 6  & 2 & {\rm Triple}     & 1.34$^t$, 0.92$^t$,
0.79$^t$, 0.50, 0.04, 0.02\\
013 & 0.10 & 3.18 & 10 & 3 & {\rm Quadruple\,\,\&\,\,binary} & 0.77$^q$,
0.68$^q$, 0.61$^q$, 0.60$^q$, 0.11$^b$, 0.11$^b$, 0.10, 0.05, 0.04,
0.06 \\
014 & 0.10 & 3.29 & 4  & 1 & {\rm Binary}     & 1.58$^b$, 1.16$^b$, 0.49,
0.08 \\
015 & 0.10 & 2.48 & 1  & 0 & {\rm Single}     & 2.48 \\
016 & 0.10 & 3.58 & 4  & 0 & {\rm Triple}     & 1.23$^t$, 1.15$^t$,
1.11$^t$, 0.09 \\
017 & 0.10 & 3.41 & 8  & 0 & {\rm Quintuple}  & 1.10$^q$, 0.98$^q$,
0.32$^q$, 0.27$^q$, 0.27$^q$, 0.15, 0.14, 0.17 \\
018 & 0.10 & 3.48 & 4  & 0 & {\rm Quadruple}  & 0.98$^q$, 0.94$^q$,
0.79$^q$, 0.77$^q$ \\
019 & 0.10 & 3.58 & 5  & 1 & {\rm Triple}     & 1.38$^t$, 1.03$^t$,
1.00$^t$, 0.11, 0.06 \\
020 & 0.10 & 3.77 & 3  & 0 & {\rm Triple}     & 1.28$^t$, 1.27$^t$, 1.22$^t$\\

041 & 0.25 & 3.13 & 6  & 0 & {\rm Quadruple}  & 0.79$^q$, 0.71$^q$,
0.68$^q$, 0.36$^q$, 0.33, 0.27 \\
042 & 0.25 & 3.14 & 8  & 2 & {\rm Triple\,\,\&\,\,binary} & 1.92$^t$,
0.41$^t$, 0.30$^t$, 0.10$^b$, 0.18$^b$, 0.07, 0.14, 0.02\\
043 & 0.25 & 2.69 & 5  & 1 & {\rm Triple}     & 0.81$^t$, 0.81$^t$,
0.53$^t$, 0.50, 0.03 \\
044 & 0.25 & 3.17 & 5  & 0 & {\rm Quadruple}  & 1.26$^q$, 0.92$^q$,
0.46$^q$, 0.27$^q$, 0.26 \\
045 & 0.25 & 3.18 & 8  & 2 & {\rm Quadruple}  & 0.78$^q$, 0.55$^q$,
0.54$^q$, 0.47$^q$, 0.05, 0.04, 0.62, 0.13 \\
046 & 0.25 & 3.19 & 3  & 0 & {\rm Binary}     & 1.70$^b$, 0.94$^b$, 0.54 \\
047 & 0.25 & 3.31 & 5  & 0 & {\rm Quadruple}  & 1.07$^q$, 0.81$^q$,
0.64$^q$, 0.46$^q$, 0.33 \\
048 & 0.25 & 3.36 & 10 & 4 & {\rm Quadruple}  & 0.87$^q$, 0.76$^q$,
0.56$^q$, 0.56$^q$, 0.08, 0.03, 0.02, 0.06, 0.37, 0.10 \\
049 & 0.25 & 2.84 & 3  & 0 & {\rm Triple}     & 2.01$^t$, 0.42$^t$, 0.41$^t$\\
050 & 0.25 & 3.37 & 7  & 2 & {\rm Quadruple}  & 0.91$^q$, 0.91$^q$,
0.68$^q$, 0.67$^q$, 0.02, 0.06, 0.12 \\
\hline
\end{tabular}
\end{table*}

\begin{figure*}
\centerline{\psfig{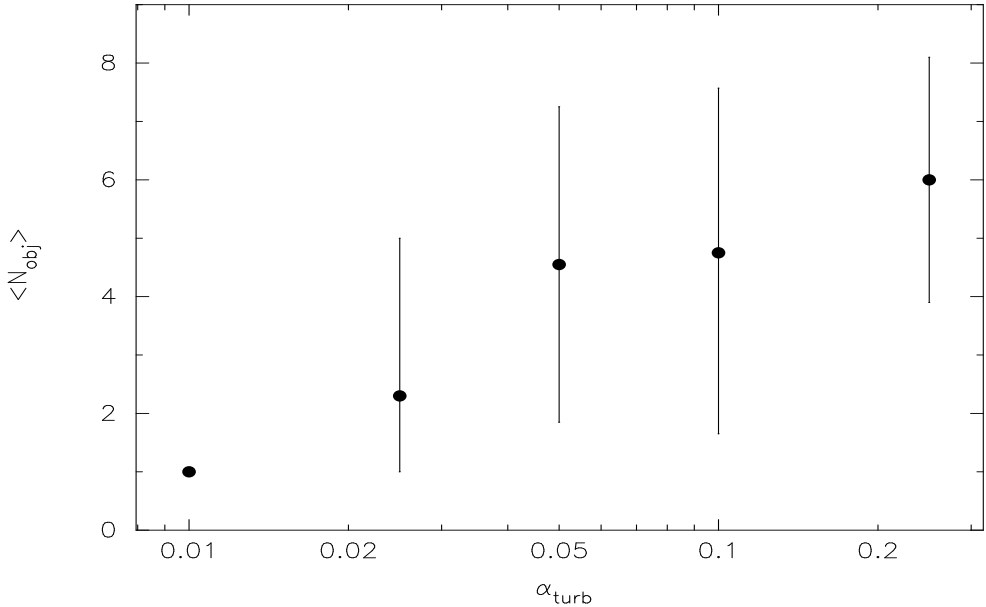}}
\caption{The average number of objects formed in a core, 
$\left<{\cal N}_{\rm obj}\right>$, as a function of the ratio of turbulent 
to gravitational energy $\alpha_{\rm turb}$. The error bars give the 
standard deviation (see Eqn. (\ref{STANDEV}).}
\label{fig:naverage}
\end{figure*}

\section{Statistics}

Details of the numbers, masses and multiplicities of the objects
produced in each of the fifty simulations with $\alpha_{\rm turb} = 
0.05,\,0.10\;{\rm and}\;0.25$ are shown in Table~\ref{tab:bins}. The 
simulations with low turbulence ($\alpha_{\rm turb} \leq 0.025$) are 
omitted because almost all of them produce only a single object, and 
therefore the discussion will now concentrate on the simulations with 
$\alpha_{\rm turb} = 0.05,\,0.10\;{\rm and}\;0.25\,$.

\subsection{Numbers of objects and formation timescale}

Increasing the level of turbulence has several effects. In the first 
instance,  it delays somewhat the time at which objects are formed, and 
at the same time it increases the average number of objects formed.

With $\alpha_{\rm turb} = 0.05$, the first object (hereafter 
the primary protostar) forms $0.05\,{\rm Myr}$ to $0.06\,{\rm Myr}$ 
after the start of the simulation, and $\sim 95\%$ of all the other 
objects have formed by $0.12\,{\rm Myr}$. With $\alpha_{\rm turb} = 
0.25$, the primary protostar forms $0.06\,{\rm Myr}$ to $0.08\,{\rm Myr}$ 
after the start of the simulation, and objects continue forming up to 
$0.15\,{\rm Myr}$; this is because the extra turbulent energy gives 
the core extra support, and therefore delays its collapse.

The majority of secondary objects form in a dense disc-like slab 
around the primary protostar. The instabilities which produce these 
secondary objects are usually seeded -- and propelled into the 
non-linear condensation regime -- by the inhomogeneities in the 
accretion flow onto the slab. Higher levels of turbulence result 
in larger inhomogeneities, and hence in more secondary objects.

For $\alpha_{\rm turb} \leq 0.01$ the mean number of objects formed 
is $\left<{\cal N}_{\rm obj}\right> = 1$ (i.e. there are no secondary 
objects), whereas for $\alpha_{\rm turb} = 0.25\,$, $\,\left<{\cal N}
_{\rm obj}\right> = 6$ (i.e. there are on average five secondary 
objects). Values of $\alpha_{\rm turb}$ and 
$\left<{\cal N}_{\rm obj}\right>$ are listed in Table 1, and plotted 
on Fig. ~\ref{fig:naverage}, where the error bars give the standard 
deviation, i.e.
\begin{equation} \label{STANDEV}
\hspace{3cm} \pm\left[\left<{\cal N}_{\rm obj}^2\right>
-\left<{\cal N}_{\rm obj}\right>^2\right]^{1/2} \,.
\end{equation}

\begin{figure*}
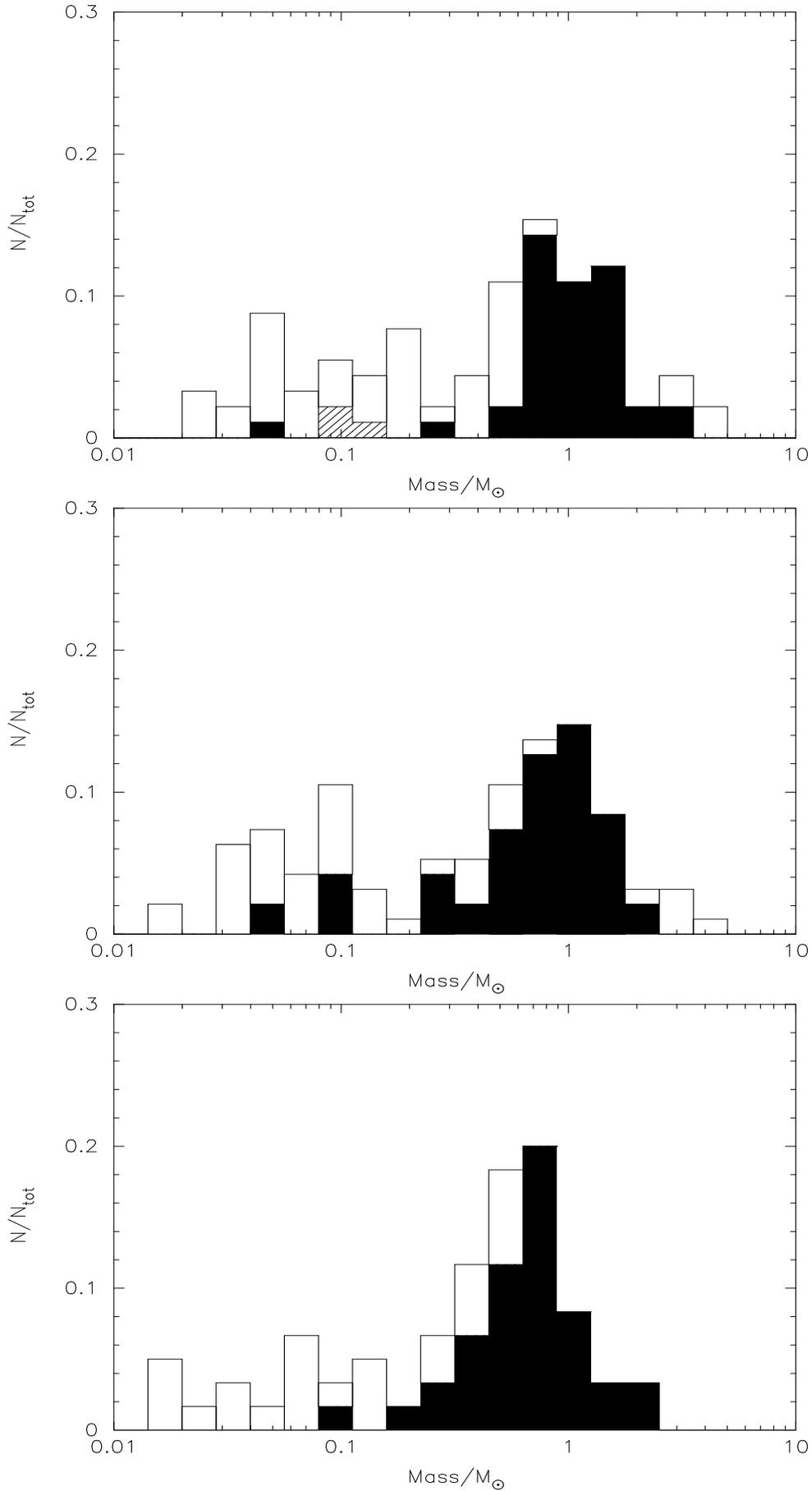

\centerline{\psfig{figure=0285fig3a.ps,height=8.0cm,width=13.0cm,angle=270}}
\centerline{\psfig{figure=0285fig3b.ps,height=8.0cm,width=13.0cm,angle=270}}
\centerline{\psfig{figure=0285fig3c.ps,height=8.0cm,width=13.0cm,angle=270}}
\caption{The normalized mass functions for $\alpha_{\rm turb} = 0.05\,
{\rm (top)},\,0.10\,{\rm (middle)\;and}\;0.25\,{\rm (bottom)}$. The 
shaded regions show stars in stable multiple systems, the hashed 
regions show stars in unstable multiple systems, and the open regions 
show single stars.}
\label{fig:imfs}
\end{figure*}

\begin{figure*}
\centerline{\psfig{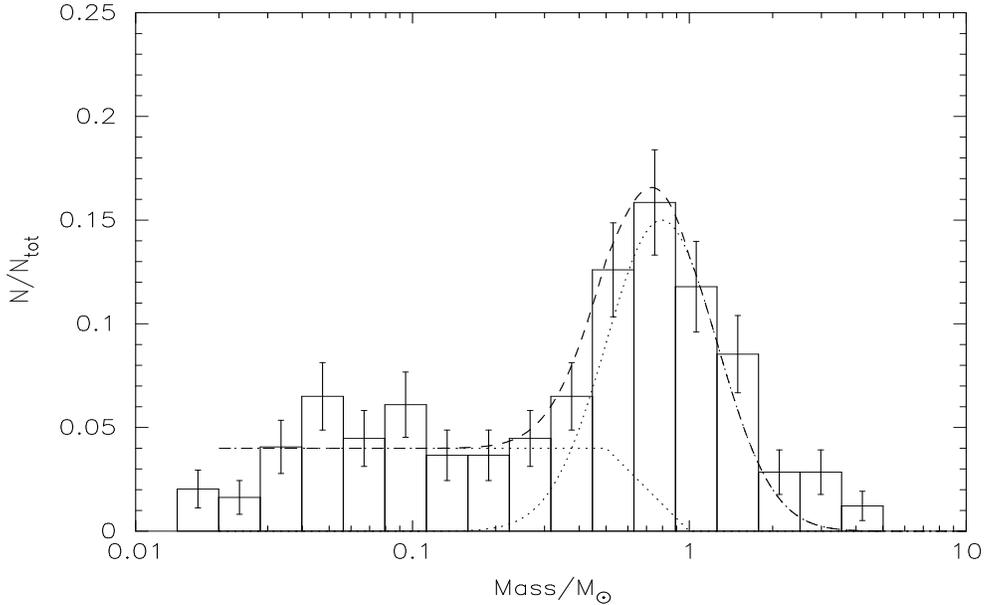}}
\caption{The combined (and un-normalized) mass function for all the 
simulations having $\alpha_{\rm turb} \geq 0.05$, (i.e. $\alpha_{\rm 
turb} = 0.05\,(\times 20),\,0.10\,(\times 20)\;{\rm and}\;0.25\,(\times 
10)$), this being an approximation to the distribution of $\alpha_{\rm 
turb}$ values in the Taurus star formation region. The error bars give 
$\pm\sqrt{N}$ uncertainties. The dotted lines show separate fits to the 
unbound objects (the flat distribution at low mass) and the bound 
objects (the log-normal distribution at high mass); parameter fits are 
given in the main text). The dashed line shows the combination of these 
two separate fits.}
\label{fig:imffit}
\end{figure*}

\subsection{The mass function}

The normalized mass functions (MF) for the ensembles with $\alpha_{\rm
turb} = 0.05,\,0.10\;{\rm and}\;0.25$ are shown in Fig.~\ref{fig:imfs}. 
In each case the shaded region shows the MF of objects in multiple 
systems and the open region shows the MF of single objects. The hashed 
region in the top figure shows an unstable triple that formed late on in 
one simulation, and is therefore likely to decay into a binary and an 
ejected single (see Paper I for more details).

The MFs are clearly very similar for each value of $\alpha_{\rm turb}$, 
{\it viz.} a high-mass peak of predominantly multiple stars and a low-mass 
tail of ejected stars and brown dwarfs. However, as $\alpha_{\rm turb}$ 
increases, 
the MF shifts slightly to lower masses. There are two reasons for this. 
(a) For higher $\alpha_{\rm turb}$ the overall collapse is delayed by 
the extra turbulent support, and therefore when the simulations are 
terminated at $0.3\,{\rm Myr}$ less mass has been incorporated into 
objects (the mean  mass incorporated into objects is given in the third 
column of Table 1). (b) For higher $\alpha_{\rm turb}$, the accretion 
flow onto the disc around the primary protostar is lumpier, so more 
objects are formed but individually they are less massive.

The combined mass function for all simulations 
with $\alpha_{\rm turb} \geq 0.05$ is shown in Fig.~\ref{fig:imffit}. 
The distribution of high-mass (predominantly bound) stars is well 
fitted by a log-normal distribution having mean 
$\left<\ell og_{10}[M]\right> = 0.05$ and standard deviation 
$\sigma_{\ell og_{10}[M]} = 0.04$. The distribution of low-mass 
(predominantly unbound) objects is consistent with being flat in 
log-space from our resolution limit at $\sim 0.025 M_\odot$ up to 
$\sim 0.5 M_{\odot}$, above which it declines. The high-mass 
(predominantly bound) stars have an average mass of $\sim 1 
M_{\odot}$ because, after ejections have removed some objects, there 
are usually two to four stars left in the core, and they are then able 
to accrete a total 
of $\sim 3 M_{\odot}$ between them. A more massive core would spawn 
more massive stars (Goodwin et al. in preparation).

The fraction of objects which are brown dwarfs is ${\cal N}_{\rm BD} 
/ {\cal N}_{\rm obj} \sim 18\%$, and there does not appear to be 
a systematic dependence on the level of turbulence. This is 
somewhat higher than in Taurus (${\cal N}_{\rm BD}/{\cal N}_{\rm obj} 
\sim 13 \%$; Brice\~no et al. 2002), and somewhat lower than in Orion 
(${\cal N}_{\rm BD}/{\cal N}_{\rm obj}\sim 26 \%$; Muench et al. 2002). 
The fraction of low-mass objects ($M < 0.5 M_\odot$) also appears to 
be independent of $\alpha_{\rm turb}$, and approximately $\sim 50\%$.

\subsection{Companion star frequencies}

In analyzing the multiplicity of the objects formed in our simulations, we 
define `systems' to include single objects, and `multiple systems' to include 
only systems containing more than one object. The primary is the most massive 
star in a system; in a single it is the only star.  Thus, if $S$ is 
the number of single objects and $B$, $T$, $Q$ and $Q'$ are the numbers 
of binary, triple, quadruple and quintuple systems, respectively, the 
total number of objects is $(S+2B+3T+4Q+5Q'+...)$, the total number of 
systems is $(S+B+T+Q+Q'+...)$ (which is the same as the total number of 
primaries), and the total number of multiple systems is $(B+T+Q+Q'+...)$.

Many different statistics have been introduced as measures of stellar 
(and brown dwarf) multiplicity (e.g. Reipurth \& Zinnecker 1993), 
and they all reflect slightly different things\footnote{Unfortunately, 
the different measures usually have more than one name. For consistency we 
have adopted the nomenclature proposed by Reipurth \& Zinnecker
1993.}. They can be divided into two groups.

The first group of measures is normalized to the total number of objects, 
and is useful because it is straightforward to derive these measures for 
a subset of objects (for example, low-mass stars, in the range $0.08 M_\odot 
< M < 0.5 M_\odot$). The companion probability (Reipurth \& Zinnecker 1993),
\begin{equation}
{\bf cp} = \frac{2B + 3T + 4Q + 5Q' + ... }
{S + 2B + 3T + 4Q + 5Q' + ... } \,,
\end{equation}
gives the probability of an object having at least one companion, 
or equivalently the fraction of objects which is in multiple systems.  
However, it gives no indication of whether objects with companions are in 
binaries, or triples, or higher multiples. Therefore we prefer the 
companion frequency, 
\begin{equation}
{\bf cf} = \frac{2B + 6T + 12Q + 20Q' + ... }
{S + 2B + 3T + 4Q + 5Q' + ... } \,,
\end{equation}
which gives the mean number of companions per object.

The second group of measures is normalized to the total number of 
systems or the total number of multiple systems. The  multiplicity 
frequency (RZ93),
\begin{equation}
{\bf mf} = \frac{B + T + Q + Q' + ... }
{S + B + T + Q + Q' + ... } \,,
\end{equation}
gives the fraction of systems which is multiple. The pairing factor 
(RZ93),
\begin{equation}
{\bf pf} = \frac{B + 2T + 3Q + 4Q' + ... }
{B + T + Q + Q' + ... } \,,
\end{equation}
gives the mean number of orbits per multiple system, or equivalently 
the mean number of companions per primary in multiple systems.

Table~\ref{tab:runs} records the values of all these measures for cores 
having $\alpha_{\rm turb} = 0.05,\,0.10\;{\rm and}\;0.25$. We have also 
calculated the companion frequency, {\bf cf}, separately for the low-mass 
objects ($<0.5 M_{\odot}$) and the high-mass objects ($>0.5 M_{\odot}$). 
The fraction of objects in multiple systems (i.e. the companion probability, 
{\bf cp}) increases steadily with increasing $\alpha_{\rm turb}$. The 
mean number of companions (i.e. the companion frequency, {\bf cf}) 
increases even more rapidly with increasing $\alpha_{\rm turb}$. {\bf cf} 
is always larger for high-mass objects than for low-mass objects (i.e. 
there are fewer low-mass objects in multiples than high-mass objects), 
but the rate of increase of {\bf cf} with increasing $\alpha_{\rm turb}$ 
is greater for the low-mass objects.

The smaller companion frequency for low-mass objects is due to the fact 
that low-mass objects have usually been ejected from the core before 
they could accrete much mass (that is why they have low mass), and 
ejected stars tend to be singles. 
Stars that end up in stable multiples also tend to remain in the core, 
and therefore they grow to larger masses by continuing to accrete.  

The multiplicity of low-mass stars increases with increasing 
$\alpha_{\rm turb}$, because a greater number of higher-order multiples 
is formed in cores with higher $\alpha_{\rm turb}$. For example, when 
$\alpha_{\rm turb} = 0.05$, only $10\%$ of simulations produce quadruples, 
but this fraction rises to $60\%$ for $\alpha_{\rm turb} = 0.25$. In 
higher-order multiples, the low-mass objects tend to be outlying 
members.  They are less able to accrete from the remaining 
gas, and they tend to remain low-mass.

\begin{figure*}
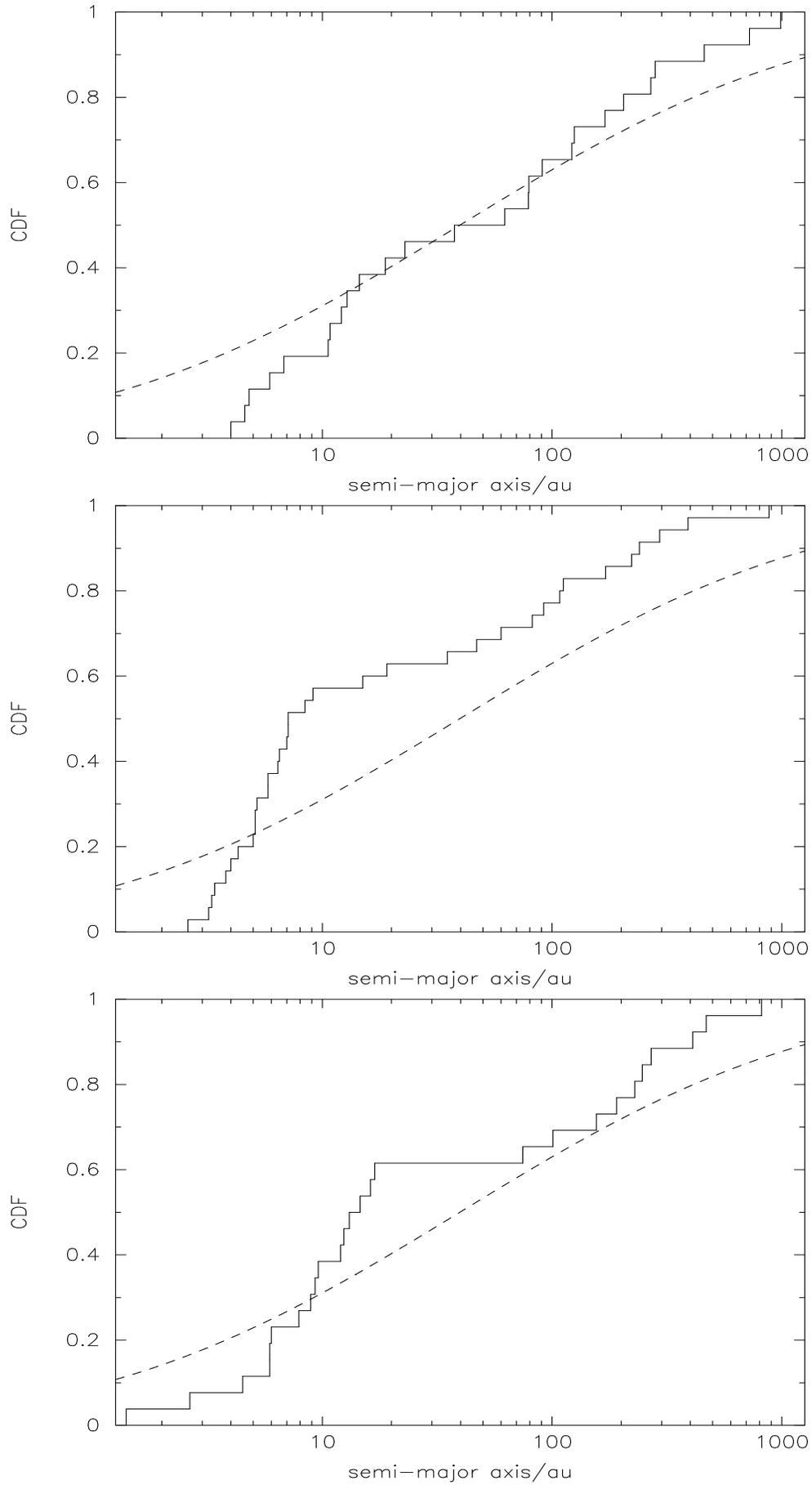

\centerline{\psfig{figure=0285fig5a.ps,height=8.0cm,width=13.0cm,angle=270}}
\centerline{\psfig{figure=0285fig5b.ps,height=8.0cm,width=13.0cm,angle=270}}
\centerline{\psfig{figure=0285fig5c.ps,height=8.0cm,width=13.0cm,angle=270}}
\caption{The histograms show the cumulative distribution functions of 
semi-major axes for the ensembles with $\alpha_{\rm turb} = 0.05\,{\rm (top)},
\,0.10\,{\rm (middle)}\,{\rm and}\,0.25\,{\rm (bottom)}$. The dashed line 
shows the guassian fit to the DM91 period distribution.}
\label{fig:semis}
\end{figure*}

\begin{figure*}
\centerline{\psfig{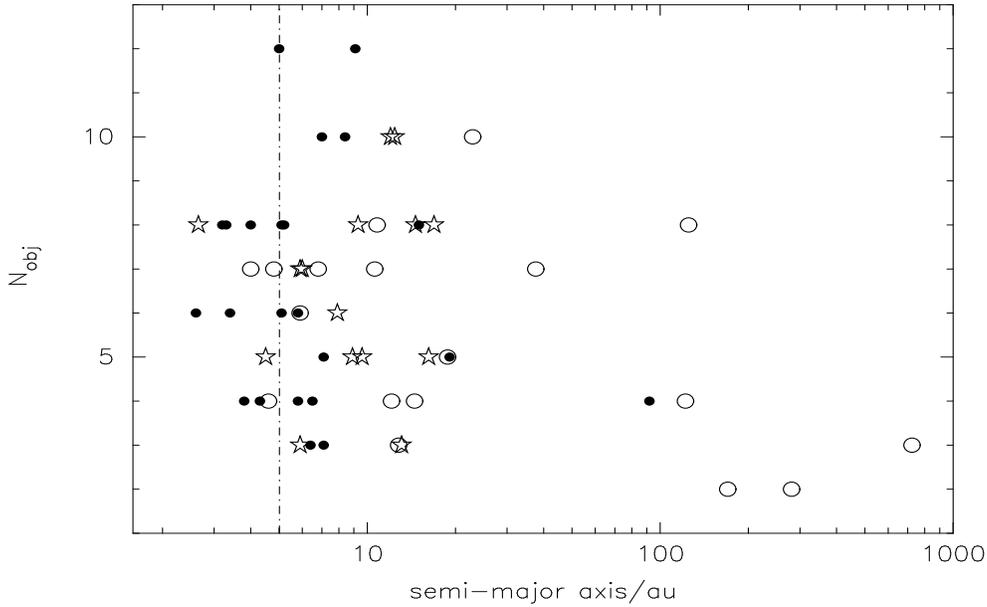}}
\caption{For each binary system, the semi-major axis, $a$ is plotted against 
the number of objects formed in that simulation, $N_{\rm obj}$, with open 
circles for  for $\alpha_{\rm turb}=0.05$, filled circles for 
$\alpha_{\rm turb}=0.10$, and stars for $\alpha_{\rm turb}=0.25$. 
The average semi-major axis, $\bar{a}$, decreases with increasing 
$N_{\rm obj}$, due to dynamical hardening. To the left of the dashed line 
at $a = 5\,{\rm au}$ the $a$-values are upper limits, due to gravity 
softening.} 
\label{fig:allnobj}
\end{figure*}

\subsection{Separations}

Fig.~\ref{fig:semis} shows the cumulative distribution functions 
(CDFs) of the semi-major axes of multiple systems, for different 
initial levels of turbulence, $\alpha_{\rm turb} = 0.05,\;0.10\;
{\rm and}\;0.25$. For comparison the gaussian fit to the DM91 
local G-dwarf sample is plotted as a dashed line.

As noted in Paper I, the semi-major axis distribution for the 
$\alpha_{\rm turb} = 0.05$ ensemble is consistent with the DM91 
observations. In contrast, the semi-major axis distributions for 
the $\alpha_{\rm turb} = 0.10\;{\rm and}\;0.25$ ensembles both have 
too many hard binaries ($a < 20$ au), and both are rejected by the KS 
test as being drawn from the DM91 fit, at $>90\%$ confidence. A similar 
excess of hard binaries is predicted by the core fragmentation simulations 
of Delgado-Donate et al. (2003, 2004), which invoke even higher 
levels of turbulence ($\alpha_{\rm turb} = 1$), and by the $N$-body 
simulations of Sterzik \& Durisen (2003).

As described in Paper I, hard binaries are formed primarily by
few-body interactions, including those which eject low mass objects. 
Consequently, in simulations where larger numbers of objects are formed, 
the binaries are on average harder (regardless of $\alpha_{\rm turb}$). 
For example, when few objects are formed, say $N_{\rm obj} \leq 3$, the 
average separation of binaries is $>100\,{\rm au}$, whereas when 
$N_{\rm obj} = 4$ the average separation is $\sim 30\,{\rm au}$, and 
when $N_{\rm obj} \geq 5$ it is $\sim 20\,{\rm au}$. Fig.~\ref{fig:allnobj} 
shows the semi-major axes of all systems plotted against the number of objects 
formed in that simulation. There is clearly a trend of decreasing 
semi-major axis with increasing number of objects.

Since the gravitational forces between objects are kernel softened with 
a smothing length $h$ equal to the sink radius $R_{\rm sink} = 5\,{\rm au}$, 
orbits with small semi-major axes ($a < 5\,{\rm au}$) are also softened. It 
follows that the distribution of semi-major axes below $5\,{\rm au}$ is 
distorted. Given that the code conserves angular momentum very accurately, 
we infer that these already hard orbits should be even harder. For 
$\alpha_{\rm turb} = 0.10\;{\rm and}\;0.25$ this would exacerbate the 
difference between the numerically derived distribution of semi-major 
axes and the observations of DM91. Conversely, for $\alpha_{\rm turb} 
= 0.05$, it would improve the agreement with the DM91 distribution.

\begin{figure*}
\centerline{\psfig{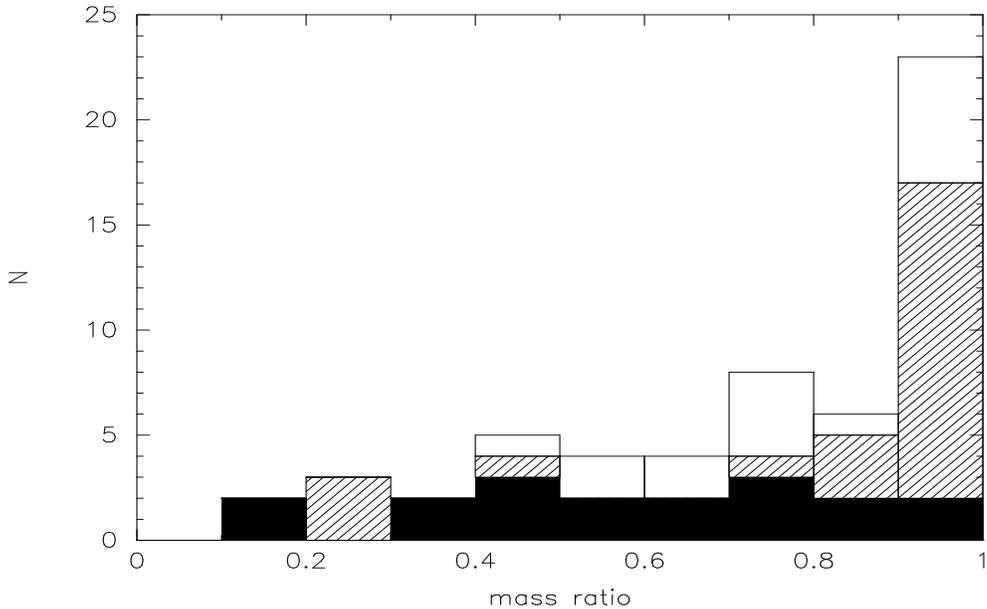}}
\caption{The distribution of mass ratios for simulations with
$\alpha_{\rm turb}=0.05$ (filled), $\alpha_{\rm turb} = 0.10$ (hashed) 
and $\alpha_{\rm turb} = 0.25$ (open).  (Note that the number of simulations 
in the $\alpha_{\rm turb} = 0.25$ ensemble is only half of the number in 
the other ensembles.)}
\label{fig:massrat}
\end{figure*}

\subsection{Mass ratios}

Fig.~\ref{fig:massrat} shows the distribution of binary mass ratios, 
$q = M_2 / M_1$, for all simulations having $\alpha_{\rm turb} \geq 0.05$. 
Note that high $q$ means $q \sim 1$, i.e. components of comparable mass.

For $\alpha_{\rm turb} = 0.05$, the distribution of mass ratios is quite 
flat, and reminiscent of the observed distribution for local G dwarfs 
(DM91; Mazeh et al. 1992).

For $\alpha = 0.10\;{\rm and}\;0.25$, the distribution is dominated by 
high-$q$ close binaries. All binaries in these ensembles have
semi-major axes $a < 100\,{\rm au}$ (the high-$a$ tail is produced by
wider orbits in higher-order systems), 
and $64\%$ of these have $q > 0.8$. This is very similar to the  mass 
ratio distribution observed in Taurus-Auriga by WG01, who 
found that over $\sim 60\%$ of binaries with separations $< 100\,{\rm au}$ 
had $q > 0.8$. In our simulations, systems with high mass ratio tend to be 
close (all systems 
with $q > 0.7$, and most systems with $q > 0.4$, are binaries with $a < 20\,
{\rm au}$), but the reverse is not always true: in other words, there are a 
few close binaries with low mass ratios. Close binaries are 
presumed to acquire high mass ratios because the material accreting onto the 
system has high specific angular momentum, and is therefore more readily 
accommodated by the secondary (Whitworth et al. 1995; Bate \& Bonnell 1997; 
Paper I).

\begin{figure*}
\centerline{\psfig{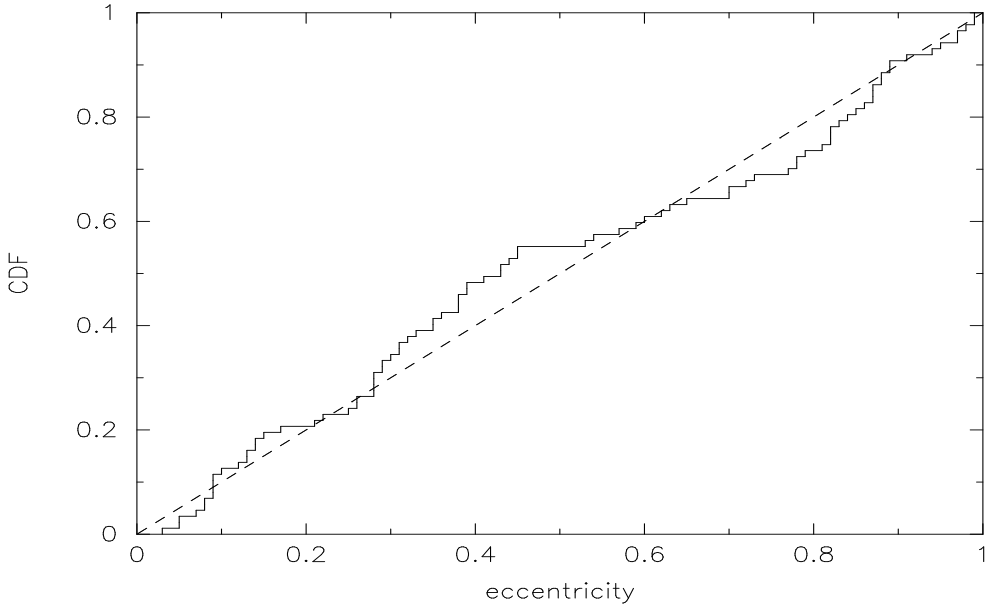}}
\caption{The cumulative distribution function of eccentricities for 
all simulations with $\alpha_{\rm turb} \geq 0.05$.  The dashed line 
shows the DM91 fit of a linearly increasing eccentricity distribution.}
\label{fig:alleccen}
\end{figure*}


\subsection{Eccentricities}

Fig.~\ref{fig:alleccen} shows the CDF of eccentricity for all the simulations 
having $\alpha_{\rm turb} \geq 0.05$ and the linear fit to the observed 
distribution proposed by DM91. The two distributions are consistent.

\section{Discussion}

\subsection{Minimum level of turbulence for multiple formation}

When the initial level of turbulence is low, $\alpha_{\rm turb} \leq 0.01$, 
it seems that a core can only spawn a single central star. Even for 
$\alpha_{\rm turb} = 0.025$, the core is unlikely to spawn a multiple system. 
We therefore focus our discussion on the higher levels of turbulence, 
$\alpha_{\rm turb} = 0.05,\,0.10\;{\rm and}\;0.25$, for which multiple star 
formation is the norm. In this range a number of significant systematic 
trends are evident.

\subsection{Time-scale for star formation}

As $\alpha_{\rm turb}$ increases from 
$0.10$ to $0.25$, the average timescale for star formation increases 
somewhat, due to the extra support which turbulence affords the core. 
For $\alpha_{\rm turb} = 0.05$, the primary protostar forms after 
$0.05\,{\rm Myr}$, and most of the secondary protostars have formed by 
$0.12\,{\rm Myr}$. For $\alpha_{\rm turb} = 0.10\;{\rm and}\;0.25$, the 
primary protostar forms after $0.06\,{\rm Myr}$, and most of the secondary 
protostars have formed by $0.15\,{\rm Myr}$.  In only one case
(run 073) do objects form after 0.25 Myr, and so it appears that the
fragmentation phase is almost always over by the end of the
simulations at 0.3 Myr.  After the end of the simulations accretion
will be on-going.  However, feedback from the protostars is expected to
become very important, possibly dispersing a significant fraction of
the gas not already in stars or discs around them.

\subsection{Number of objects formed}

As $\alpha_{\rm turb}$ increases from $0.10$ 
to $0.25$, the average number of objects formed, $\left< {\cal N}_{\rm obj} 
\right>$, increases from $4.6 \pm 2.6$ to $6.0 \pm 2.0$ (see 
Fig.~\ref{fig:naverage}). This is because a higher level of turbulence 
generates more density contrast -- i.e. more numerous and more compressed 
lumps -- and therefore more protostars.

\subsection{Masses of objects formed}

As $\alpha_{\rm turb}$ increases from $0.10$ 
to $0.25$, the average mass of the objects formed decreases slightly (see 
Fig.~\ref{fig:imfs}). There are two factors involved here. First, as noted 
above a higher level of turbulence means that more objects are formed. Second, 
a higher level of turbulence means that the core has more support, and 
therefore a smaller fraction of its mass has condensed out after $0.3\,{\rm Myr}$.

\subsection{The bimodal mass function}

Apart from this slight decrease in average 
mass with increasing 
turbulence, the form of the mass function appears to be independent of 
$\alpha_{\rm turb}$. Specifically, the mass function is bimodal: the 
lower-mass stars (which tend to be single stars ejected from the core) 
subscribe to a flat segment of the mass function; and the higher-mass stars 
(which tend to be those remaining in the core and pairing up in multiple 
systems) subscribe to a Gaussian segment of the mass function (see Fig.
~\ref{fig:imffit}).

The critical mass seperating the two modes in the mass function,  
$M_{\rm crit} \sim 0.5 M_\odot$, arises because of the interplay 
between ejection by dynamical interaction and growth by accretion. 
Anosova (1986) has shown that the decay time for small-$N$ systems is 
\begin{equation}
t_{\rm decay} \sim 100\,t_{\rm cross} \sim 17\,\,
\left( \frac{R}{\rm au} \right)^{3/2}\,
\left( \frac{M}{M_\odot} \right)^{-1/2}\,\,{\rm yr},
\end{equation}
and the ensemble of low-mass secondaries which forms in our cores 
typically has $R \sim 200\,{\rm au}$ and $M \sim 1\;{\rm to}\;2 M_\odot$, 
so $t_{\rm decay} \sim 3\;{\rm to}\;5 \times 10^4\,{\rm yr}$. The 
accretion rate is $\sim 10^{-5} M_\odot\,{\rm yr}^{-1}$, and so the 
objects which get ejected have masses $\la 0.5 M_\odot$. The 
probability of ejection is only weakly dependent on mass, $\propto 
M^{-1/3}$ (Anosova, 1986), and so the ejected objects have a flat 
mass function. In contrast, the two or three objects which survive 
the dynamical decay phase remain embedded in the centre of the core 
and compete for the gas which continues to fall into the centre, so 
they grow to $\sim M_\odot$.

\subsection{Companion star frequency}

As $\alpha_{\rm turb}$ increases from 
$0.10$ to $0.25$, the 
companion star frequency increases slightly for intermediate-mass stars 
($0.5 M_\odot$ to $5 M_\odot$), and quite markedly for low-mass objects 
($< 0.5 M_\odot$). 

\subsection{Distribution of semi-major axes}

For $\alpha_{\rm turb} = 0.05$, the distribution of semi-major 
axes is broad and indistinguishable from the distribution inferred for 
local G-dwarfs by DM91. There is a lack of very close systems 
with $a < 5\,{\rm au}$. This is due to the fact that sinks have a finite 
size and their mutual gravity is softened; therefore our code cannot 
resolve very close systems.

In contrast, for $\alpha_{\rm turb} = 0.10\;
{\rm and}\;0.25$, there are many more close systems ($5\,{\rm au} \la 
a \la 20\,{\rm au}$) than in the DM91 sample (see Fig.~\ref{fig:semis}), 
and this discrepancy would not be alleviated if the code were able to 
resolve very close systems.

Much of the hardening which produces close binaries is due to dynamical 
interactions with other objects, in particular with the low-mass objects which 
get ejected in the process. The excess of close systems produced 
by higher levels of turbulence may therefore be due to the greater number 
of objects formed, and hence the greater potential for dynamical 
interactions, as suggested by Fig.~\ref{fig:allnobj}.

\subsection{Distribution of mass-ratios}

For $\alpha_{\rm turb} = 0.05$, the distribution of mass ratios 
is flat and indistinguishable from the distribution for local G-dwarfs 
reported by DM91. In contrast, for $\alpha_{\rm turb} = 0.10\;{\rm 
and}\;0.25$, there is an excess of systems having high mass-ratio, i.e. 
components of comparable mass (see Fig.~\ref{fig:massrat}). Many of these 
systems with high mass-ratio arise because the binary system has accreted 
material with relatively high specific angular momentum, and this 
material can more easily be accommodated by the secondary (e.g. 
Whitworth et al. 1995).

\subsection{Close systems with comparable components}

For $\alpha_{\rm turb} 
= 0.10\;{\rm and}\;0.25$, the systems with high mass-ratio tend also to 
be close, and it is this sub-population of high--mass-ratio close binaries 
which is the main difference between the distributions of semi-major axis 
and mass-ratio for the protostars formed in these simulations, and the 
distributions of semi-major axis and mass-ratio for local G-dwarfs as reported 
by DM91. A similar excess of close systems with comparable components was found 
by Delgado-Donate et al. (2003, 2004), who simulated the collapse and 
fragmentation of cores with even higher levels of turbulence ($\alpha_
{\rm turb} = 1$).

Taken at face value, this suggests that the local 
population of G-dwarfs must have been formed in cores with low 
turbulence ($\alpha_{\rm turb} \sim 0.05$). However, this conclusion 
rests on the assumption that the spherically symmetric $5.4 M_\odot\,$ 
core and the $P_k \propto k^{-4}$ turbulence spectrum which we have 
adopted, are representative of the cores forming G-dwarfs, and there 
is no firm basis for this assumption. 
An alternative explanation is that a significant population of close, 
high--mass-ratio systems has escaped detection, but we believe this to 
be unlikely. 
 
A significant contrast to this is found in Taurus. Here WG01 find that
 binaries in the separation range 
$10\,{\rm au} \la a \la 100\,{\rm au}$ do indeed have significantly higher mass 
ratios than wider binaries. Therefore they are compatible with formation 
in cores having higher levels of turbulence, $\alpha_{\rm turb} = 0.10\;
{\rm to}\;0.25$. We have discussed the origin of the mass function and the 
binary statistics in Taurus in Goodwin et al. (2004b).

\section{Conclusions}

We have explored the influence of turbulence on the fragmentation of 
dense molecular cores, by means of a large ensemble of simulations. 
In this ensemble, we consider a spherically symmetric $5.4 M_\odot$ 
core with a Plummer-like density profile; 
this is a good representation of observed cores like L1544. We seed the 
core with a turbulent velocity field having power spectrum $P(k) 
\propto k^{-4}$. The number of objects that forms, and the properties 
of the resulting multiple systems depend both on the level of turbulence 
$\alpha_{\rm turb}$, and on the details of the turbulent velocity 
field. Therefore for each value of $\alpha_{\rm turb}$ we we have simulated 
many different realizations by changing the random number seed for the 
turbulent velocity field. The main conclusions are

(i) The formation of multiple systems requires $\alpha_{\rm turb} \ga 
0.025$; a core with $\alpha_{\rm turb} = 0.025$ has a $\sim 20\%$ chance 
of forming a multiple system, and a core with $\alpha_{\rm turb} \ga 0.25$ 
almost always forms a multiple system.

(ii) As $\alpha_{\rm turb}$ is increased, the average time-scale for 
object formation increases, the average number of objects formed increases, 
the companion frequency increases (particularly for the lower-mass objects), 
and the average mass of objects decreases.

(iii) The mass function has a bimodal form. The low-mass objects, which 
are usually single because they have been ejected from the core before 
they could grow above $0.5 M_\odot$, subscribe to a flat segment of the 
mass function. The high-mass stars, which have usually  stayed embedded 
in the core, grown by accretion and paired up with one another 
in multiple systems, subscribe to a Gaussian segment of the mass 
function. Typically $20\%$ of objects are brown dwarfs ($M < 0.08 M_\odot$), 
and $50\%$ are low-mass stars ($0.08 M_\odot < M < 0.5 M_\odot$).

(iv) For $\alpha_{\rm turb} \ga 0.10$, there is a significant subpopulation 
of binary systems having small semi-major axes and high mass ratios, i.e. close 
systems with components of comparable mass. This subpopulation is also found 
in the simulations of Delgado-Donate et al. (2003, 2004) who treat the extreme 
case $\alpha_{\rm turb} = 1$. It is not present in the sample of local G-dwarfs 
observed by DM91, but there is some evidence for it in the Taurus 
pre--Main-Sequence sample observed by WG01.

(v) The ensemble of simulations for cores with $\alpha_{\rm turb} = 0.05$ 
reproduces the binary statistics of the DM91 sample (companion-star frequency 
and distributions of 
semi-major axis, eccentricity and mass-ratio) very well. Therefore if the 
other core parameters, which we have not varied (e.g. mass), are representative 
of the cores forming local G dwarfs, we infer that these cores must have had 
finite but low levels of turbulence, $\alpha_{\rm turb} \sim 0.05$.

(vi) Both the mass function, and the binary statistics, for the WG01 
sample of pre-MS stars in Taurus are reproduced by a mix of simulations 
with $\alpha_{\rm turb} =$ 0.05 (20 realizations), 0.10 (20 realizations), 
and 0.25 (10 realizations), as shown by Goodwin et al. (2004b).

\begin{acknowledgements}

SPG acknowledges support of PPARC grant PPA/G/S/1998/00623 and is now
a UKAFF Fellow.  We thank B. Sathyprakash and R. Balasubramanian for 
allowing us extensive use of the Beowulf cluster of the gravitational 
waves group at Cardiff; and Matthew Bate for helpful discussions and 
for providing the code to generate turbulence.  

\end{acknowledgements}

\end{document}